\documentclass[
 aps, pra,
 amsmath,amssymb,
 notitlepage,
 twocolumn,
superscriptaddress,
nofootinbib ]
{revtex4-2}

\usepackage{graphicx}
\usepackage{dcolumn}
\usepackage{bm}
\usepackage{braket}
\usepackage{comment}

\usepackage{amsmath, amsthm, amssymb,mathrsfs}

\usepackage{todonotes}

\begin{document}

\title{Bell Inequality Violations Without Entanglement?\\ It's Just Postselection}

\author{Ken Wharton}
\affiliation{Department of Physics and Astronomy, San Jos\'e State University, San Jos\'e, CA 95192-0106}
\author{Huw Price}
\affiliation{Trinity College, Cambridge CB2 1TQ, UK.}

\begin{abstract}
Recently Wang \textit{et al} have reported a violation of a Bell inequality without entanglement. We show that their result is an artifact of postselection. It is well known that postselection may yield Bell inequality violations, both in classical toy models and in real experiments with delayed-choice entanglement-swapping. Here we describe a classical analog of Wang \textit{et al}'s experiment, and show that it produces essentially the same results as their quantum version. We explain in detail why neither version is a challenge to Local Causality or local realism: the postselection entails a rejection of Bell's assumption of Statistical Independence.
\end{abstract}

\maketitle

\section{Introduction}

A recent paper by Wang \textit{et al}~\cite{Wang} reports an experimental Bell inequality violation which the authors say “cannot be described by quantum entanglement in the system”.  Bell inequality violations are interesting because, under the assumptions of Bell's Theorem \cite{bell64,bell1976,bellSEP}, they cannot be achieved by a hidden variable model respecting Bell's ``Local Causality''.  This new result is surprising because it is widely assumed that measurements on separable (non-entangled) states can always be reproduced by a Locally Causal hidden variable model. Wang \textit{et al} claim to offer a counterexample.

In this paper, we show that Wang \textit{et al} likely misidentify the source of their Bell inequality violation: the observed behavior can be entirely attributed to the postselection in the experiment they describe.  We confirm this by analyzing the same experimental geometry using classical electromagnetic theory, supplemented with hidden input fields and a threshold detection rule.  Such a classical hidden variable model respects Local Causality in every way, and also allows a careful analysis of the postselection process. This process yields very similar `violations' of the same Bell inequality.

It is well known that postselection can create Bell inequality violations without entanglement. Indeed, much stronger correlations are easy to achieve. To adapt a toy model from \cite[\S4]{PriceWharton21a}, suppose that Alice and Bob each choose a binary `setting', and flip a coin to yield an `outcome'. Each sends the resulting two bits to Charlie. With repeated runs of this procedure, Charlie can discard results selectively, to yield Bell inequality violating correlations, or even a perfect correlation between (say) Alice's setting $\alpha$ and Bob's outcome $B$, in the subensemble of retained results. 

Clearly, this example involves no nonlocality. Even in the perfect correlation version, it cannot be used for control or signaling from Alice to Bob. Alice does not influence Bob's coin toss. Suppose $\alpha=B$, in a particular run. If Alice had chosen a different setting, would Bob's coin toss also have had a different result?  No. In that case $\alpha$ and $B$ would have differed, and Charlie would have discarded the result. In the language of causal modeling, such postselected correlations are `selection artifacts', a result of so-called `collider bias' or Berkson's bias \cite{holm22,Causality}. They are said to be \textit{counterfactually fragile} \cite{PriceWharton21a}.

In this toy example, Charlie has access to Alice and Bob's `settings' and `outcomes', and could not postselect such correlations without it. In real Bell experiments with postselection, the geometry may be chosen to exclude such access. For example, in Bell experiments based on entanglement-swapping \cite{shalm2015,giustina2015,rosenfeld2017}, a joint measurement outcome can be postselected \textit{before} the settings are chosen. The analysis is more subtle in cases of delayed-choice entanglement-swapping, where the postselected Bell inequality violation is not a result of entanglement between the measured particles at the time the measurement was made \cite{gaasbeek10,egg13,fankhauser19,guido21,PriceWharton21a, mjelva24}. Even then, it is sometimes possible to prove that a violation of Local Causality must be occurring \textit{somewhere} in the experiment \cite{hensen2015,PriceWharton21a}.

But in the experiment described in \cite{Wang}, the postselection is not protected in this manner.  Outcomes are determined by a coincidence event between four different regions, and the ``settings'' are in the past lightcones of these regions.  In this situation, collider bias serves to break one of the assumptions required to derive Bell inequalities (Statistical Independence, discussed in detail below).  Violation of Bell inequalities in such a circumstance is therefore neither surprising nor meaningful, as will be emphasized by a classical analysis of the same experimental geometry.

One useful consequence of the below analysis is to draw a sharper distinction between postselection and the ``detection efficiency loophole''\cite{sample0,sample1,sample2,sample3,sample4,sample5,la2021}. Analysis of restricted-postselection scenarios (such as entanglement swapping) might lead to the mistaken conclusion that better detectors would always make postselection effects safely ignorable.  But for the more general postselection considered here, those two issues are seen to decouple: no level of detection efficiency would change the implications of this experiment.  Again, this conclusion is borne out by the classical analysis. 

Another use of the classical analog model is to assess the technique proposed in \cite{Wang} whereby a ``complementary set'' of measurement settings aims to restore a form of Statistical Independence.  In order for this work, each complementary set would have to fairly sample the same hidden variable space.  Although there is no way to test this assumption in the actual experiment, the classical analog model reveals that their chosen sets do not restore Statistical Independence; the (postselected) initial hidden variables remain correlated with the choice of which set to use.

We also hope that our analysis will highlight the deep and often-ignored connection between classical electromagnetism and quantum optics.  The fact that a surprising quantum result can be recovered from an analysis based entirely on classical electromagnetism might be interesting in its own right, even though this is hardly the first example \cite{boyer1980,zhu1986,fain1987,la2021,WhartonAdlam}.

But, most importantly, we hope to draw the quantum foundation community’s attention to the way that the Bell-inequality-violating results presented in \cite{Wang} can also be found in a classical analog of the same experiment.  Since the classical violations can easily be explained via postselection, we argue that the results seen in the original experiments can also be explained in this manner, and are not therefore indicative of inherently quantum phenomena.\footnote{We ourselves have recently argued that ordinary cases of entanglement may be regarded as selection artifacts, in cases in which the selection is a physical constraint on a suitable portion of a full history \cite{pricewharton25}. We set this more general issue aside, for the purposes of the present piece.}

\section{Classical Analysis of a Quantum Experiment}

With a few simple assumptions, it is possible to take the experimental geometry utilized in Wang \textit{et al} and analyze it with classical electromagnetism, rather than quantum optics.  This section describes this procedure, and motivates the resulting classical model, with the same outcome probabilities as the quantum experiment of \cite{Wang}. Readers not interested in the technical motivation may skip to the next section, where the classical model is summarized.

Our simplified version of the experiment from Wang \textit{et al} can be seen in Figure 1.  Four nonlinear crystals, $I,II,III$, and $IV$, are each pumped by a strong laser (thick arrows).  Classically, each pump laser is just an electromagnetic (EM) wave mode, in a nonlinear medium which couples to two other classical EM modes (thin arrows) via three-wave mixing.  In crystal $I$, in the limit that both input field modes $s_{10}$ and $p_{10}$ are exactly zero, no amplification would be observed.  In order to classically model spontaneous parametric down-conversion, it is therefore necessary to consider that these input modes are not zero.  

\begin{figure}[htbp]
\begin{center}
\includegraphics[width=8cm]{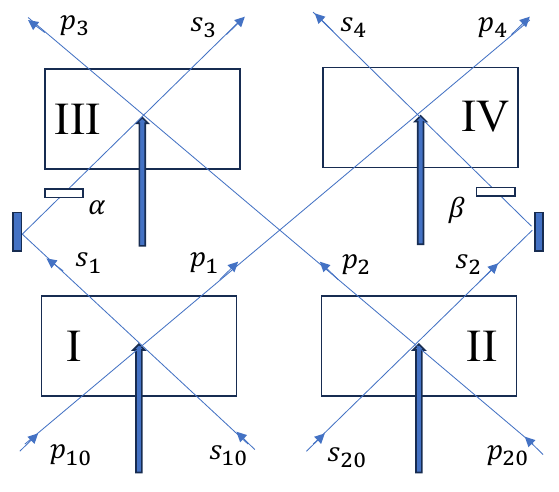}
\caption{A simplified version of the experiment described in \cite{Wang}.  Four nonlinear crystals are each pumped with a strong laser (thick arrows), each laser tuned for a resonant three-wave interaction with two other electromagnetic modes (thin arrows).  The bottom crystals are seeded from incoming classical fields with complex amplitudes $s_{10},p_{10},s_{20},$ and $p_{20}$.  The top crystals then can further amplify the outputs from the bottom.  The rectangles labelled $\alpha$ and $\beta$ are adjustable phase plates. }
\label{fig1}
\end{center}
\end{figure}

This step of populating the incoming field modes is commonly done in any semi-classical analysis of spontaneous emission processes, where it is convenient to imagine that each EM field mode is populated with a field intensity corresponding to ``half a photon’’ per mode.  We will make this same assumption.  Such a semi-classical step may make the analysis appear not strictly ``classical’’, but note that an unknown incoming field is precisely the sort of initial hidden variable considered by Bell, and the question at hand is whether or not a hidden variable model constrained by Local Causality can reproduce the results of Wang \textit{et al.} The only other hidden variable in this model is the phase of each of these incoming modes.  We will assume that these phases are independent, each randomly distributed -- another standard assumption required to recover spontaneous emission in semi-classical theories.  

In these experiments it is common that the EM waves in question are nearly monochromatic with well-defined spatial modes.  These modes are constrained by frequency-matching conditions; $\omega_{pump}=\omega_s+\omega_p$, corresponding to energy--conservation.  The classical model will work even if $\omega_s\ne\omega_p$; there is no need for ``identical photons’’ in any given crystal.

It is much easier to analyze classical three-wave mixing if the envelope of each wave can be described by a single complex field amplitude $E(t)$ (complex to encode classical phase information).  Assuming the waves can all interact as complete systems, there is no essential need to change the shape of the envelope with time or add additional spatial parameters to $E(t)$.  The only other assumption needed is that the intensity in the pump mode is much larger than the intensity in the two coupled modes, a limit that is certainly compatible with the reported experiment in \cite{Wang}.  

In this limit, following the classical analysis in \cite{WhartonAdlam}, the output fields at $t\!=\!t_m$ from crystal $I$ can be deterministically calculated from the input fields at $t\!=\!t_i$.  Defining the frequency-scaled complex input mode amplitudes $s_{10}=E_s(t_i)/\sqrt{\omega_s}$ and $p_{10}=E_p(t_i)/\sqrt{\omega_p}$, the corresponding output modes $s_{1}=E_s(t_m)/\sqrt{\omega_s}$ and $p_{1}=E_p(t_m)/\sqrt{\omega_p}$ have amplitudes of
 \begin{eqnarray}
s_{1} &=&  s_{10} \cosh(g) + i p_{10}^* \sinh(g) \label{eq:A1} \\
p_{1}^* &=&  p_{10}^* \cosh(g) - i s_{10} \sinh(g). \label{eq:A2}
 \end{eqnarray}
Here $g$ is a unitless gain parameter, summarizing the strength of the pump beam, the three-wave coupling efficiency, and other relevant factors.  We will assume $g$ is the same for each of the four crystals.

Since crystal $II$ works the same as crystal $I$, the bottom half of the experimental geometry shown in Figure 1 can be summarized by a linear transformation of the complex field amplitudes $\ket{F(t_m)}=\hat{G}\ket{F(t_i)}$, meaning
\begin{widetext}
\begin{equation}
\label{eq:half}
\begin{pmatrix} s_1 \\
p_1^* \\
s_2 \\
p_2^*
\end{pmatrix} 
= \begin{pmatrix} \cosh (g) & i\sinh(g) & 0 & 0 \\
-i \sinh(g) & \cosh(g) & 0 & 0 \\
0 & 0 & \cosh (g) & i\sinh(g) \\
0& 0&-i \sinh(g) & \cosh(g)
\end{pmatrix}
\begin{pmatrix} s_{10} \\
p_{10}^* \\
s_{20} \\
p_{20}^*
\end{pmatrix}.
\end{equation}
\end{widetext}

After the first amplification stage, the $s_1$ and $s_2$ modes are each sent through a phase plate (set to angles $\alpha$ and $\beta$ respectively) before seeding the next nonlinear crystals $III$ and $IV$.  Notice that these latter crystals implement \textit{stimulated} parametric down-conversion, and do not need any additional input seeds.  Also, the beams are interchanged in the middle; $s_1$ and $p_2$ are sent into crystal $III$, while $s_2$ and $p_1$ are sent into crystal $IV$.  All of these intermediate transformations are summarized by the matrix $\hat{T}$ shown below.

In the experiment, the input to these latter crystals must also be frequency-matched with the pump beam.  So if all pump beams have the same frequency $\omega_{pump}$, then it is important that all $s-$modes have the same frequency, and also that all $p-$modes have the same frequency.  (In a sense, the corresponding ``photons'' are ``identical'', but only because one crystal's output field is matched to another crystal's input.)

Implementing the rest of the experiment, from $t=t_m$ to $t=t_f$, corresponds to the linear transformation $\ket{F(t_f)}=\hat{G}\,\hat{T}\,\ket{F(t_m)}$, meaning
\begin{equation}
\label{eq:half2}
\begin{pmatrix} s_3 \\
p_3^* \\
s_4 \\
p_4^*
\end{pmatrix} 
= \hat{G} \begin{pmatrix} e^{i\alpha} & 0 & 0 & 0 \\
0 & 0 & 0 & 1 \\
0 & 0 & e^{i\beta} & 0 \\
0& 1&0 & 0 
\end{pmatrix}
\begin{pmatrix} s_{1} \\
p_{1}^* \\
s_{2} \\
p_{2}^*
\end{pmatrix},
\end{equation}
where the $\hat{G}$ matrix is the same as the one shown in (\ref{eq:half}).

Simulating the classical version of this experiment simply requires choosing a coupling parameter $g$ and starting with the classical equivalent of ``half a photon'' per mode, each with a random phase ($\delta_i$):
\begin{equation}
\label{eq:input}
\ket{F(t_i)}= 
\begin{pmatrix} \sqrt{0.5} e^{i\delta_1} \\
\sqrt{0.5} e^{i\delta_2} \\
\sqrt{0.5} e^{i\delta_3} \\
\sqrt{0.5} e^{i\delta_4}
\end{pmatrix}.
\end{equation}

The original experiment tested whether there was exactly one photon in each of the output modes $\ket{F(t_f)}$, and postselected only those results.  Although there is not a good classical analog for a precise single-photon detection in this framework, consider that we are using the field amplitude $|s_{10}|=\sqrt{0.5}$ to represent half a photon (the square root is to distinguish field amplitude from intensity/energy).  This field strength must be insufficient to trigger a photon detection; classically, there is simply not enough energy in the mode.  Since partial-photon field strengths are classically possible, the evident classical analog of this postselection process is not to impose $|s_3|=\sqrt{1.0}$ exactly, but rather to ask whether $\sqrt{1.0}\leq |s_3| < \sqrt{2.0}$.  If so, then there would be sufficient energy to allow a single-photon detection on mode $s_3$, without measuring two photons.  

With this in mind, our classical model performs an analog to the quantum measurement procedure of the original experiment by asking how often each of the four output field modes lie in the $\sqrt{1.0}$ to $\sqrt{2.0}$ range.  We can further simplify this procedure by choosing a low amplification factor $g=0.25$, making large $|s|\ge\sqrt{2.0}$ fields impossible given the small $|s|=\sqrt{0.5}$ inputs.  This calculation becomes easily tractable: a Monte Carlo simulation can choose four random phases $\delta_i$ (corresponding to Bell's initial hidden variables $\lambda$) and determine the fraction of the simulations that lead to at least $\sqrt{1.0}$ in each of the four outputs of $\ket{F(t_f)}$.  These results are shown in the next section.

\section{Model Summary and Results}

We now describe a classical hidden variable model, respecting Bell's Local Causality, motivated by the above discussion.  First, start with an input vector $\ket{F(t_i)}$ defined by Eqn. (\ref{eq:input}), comprised of four equal amplitudes with four random phases.  Then, deterministically generate an output vector using
\begin{equation}
    \ket{F(t_f)}=\hat{G}\hat{T}\hat{G}\ket{F(t_i)},
\end{equation}
with $g=0.25$ in the above definition of $\hat{G}$.  Finally, consider a ``detection'' to occur when each component of $\ket{F(t_f)}$ has a magnitude of at least $1.0$.  Clearly, it is impossible for quantum effects to be relevant to this straightforward model.

\begin{figure}[htbp]
\begin{center}
\includegraphics[width=9cm]{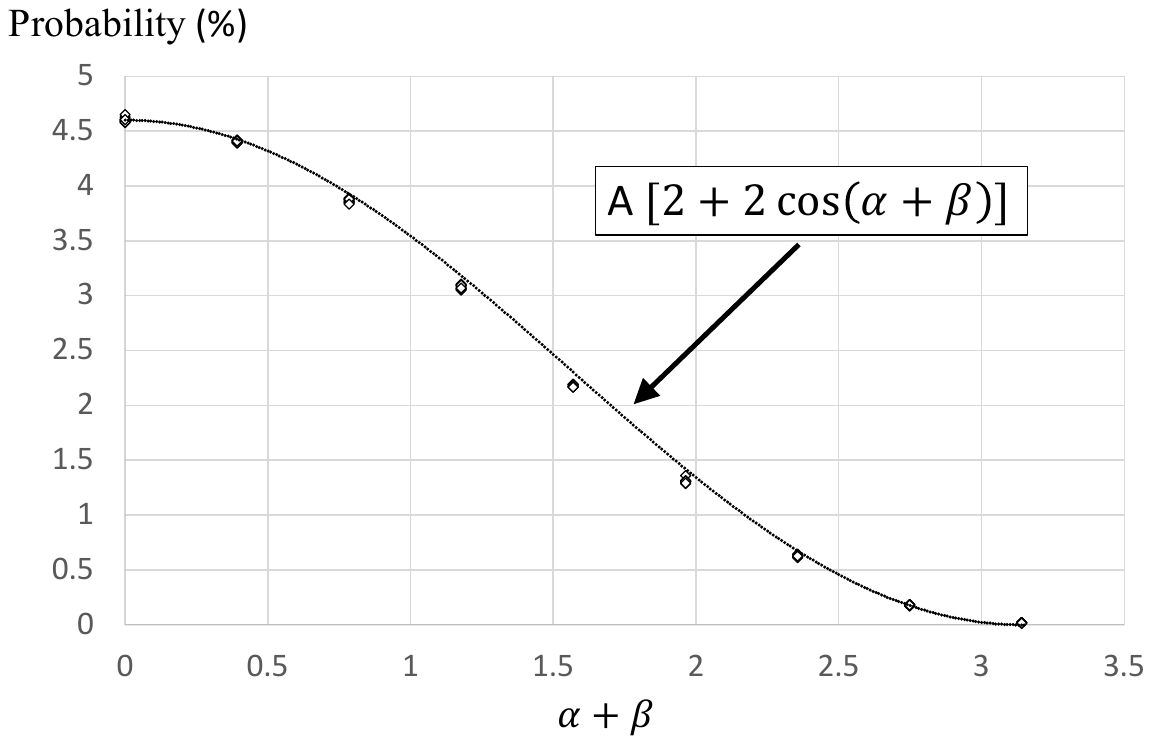}
\caption{Results from Monte Carlo simulations of the classical model; the vertical axis is the probability that a given random initial seed would lead to all four outputs exceeding the postselection threshold.  Each cluster of results represents 10 independent runs of $10^6$ seeds each.  Five of these runs use $\alpha=0$; the other five use a random value for $\alpha$ and adjust $\beta$ accordingly.  A comparison curve is shown for the value $A=0.0115$.}
\label{fig2}
\end{center}
\end{figure}

And yet, one can easily see that the same ``Bell inequality violations'' described in \cite{Wang} are also present in this calculation.  A simple computer program was used to generate many independent inputs and count the relative number of ``detections'' as defined above.  The probability of a detection was only sensitive to the quantity ($\alpha+\beta$), with results shown in Figure 2.  The probabilities lie close to a $2+2\cos(\alpha+\beta)$ curve, the same prediction as the quantum calculation.\footnote{This match turns out to be a special case; for larger values of $g$ the curve is shifted upward such that the probability does not go to zero for $(\alpha+\beta)=\pi$; for lower values of $g$ the curve is shifted downward, flattening out at zero probability.}  The next section will analyze this result.

\section{Discussion}

The first major clue as to how a classical model can violate a Bell inequality is to look at the marginal detection events on each side of the experiment (the outputs from Crystal $III$ on one hand, and crystal $IV$ on the other).  The classical calculation is easily modified to give a detection event when the two modes coming out of one single crystal both have an amplitude greater than $1.0$ (two modes, not all four).  Again, in agreement with the results reported in \cite{Wang}, these marginal probabilities are completely independent of either $\alpha$ or $\beta$.  For the gain value $g=0.25$, the marginal probability of a ``detection'' on either side of the experiment is about $20.65\%$.  This value is much larger than the peak joint detection probability of $4.6\%$ shown in Figure 2; therefore, it often occurs that a detection is made on one side, but not the other.  These cases are thrown away during the postselection process described in \cite{Wang}.

In some ways, this situation is reminiscent of the detection-efficiency loophole, where some events are (unintentionally) ``discarded'' when the detectors do not fire.  But in the standard framing of that loophole \cite{sample0,sample1,sample2,sample3,sample4,sample5,la2021}, there are three possibilities on each arm of an entanglement experiment: two possible measured outcomes, or a non-outcome.  This experiment only considers one outcome on each arm, counting a pair of photon detection events as a single outcome, with everything else treated as a non-outcome.  And since the analysis in \cite{Wang} is entirely based on relative detection rates, increasing or decreasing the detector efficiency would do nothing to change the conclusions.  Indeed, the classical model gives the same correlations as the experiment assuming a perfect ability to measure the strength of each field mode, so ``detector efficiency'' is of only tangential relevance.

Instead, the relevant issue is the winnowing of the possibility space of the experiment imposed by postselection; see Figure 3 for a schematic cartoon.  An event that only happens with (say) $4.6\%$ probability in the larger space could be said to happen with a much larger probability in a smaller, postselected subset.  And in the case of the experiment in \cite{Wang}, all that can be measured is a \textit{rate}, not a probability.  To transform this setting-dependent rate $N(\alpha,\beta)$ into a probability $p$, it must be normalized in some manner -- divided by some ``partition function'' $Z$, such that $p=N/Z$.  With a strict postselection, $Z$ could become small enough to increase the probabilities into Bell-inequality-violating territory, even if no such violations are present in the larger space.

\begin{figure}[htbp]
\begin{center}
\includegraphics[width=6cm]{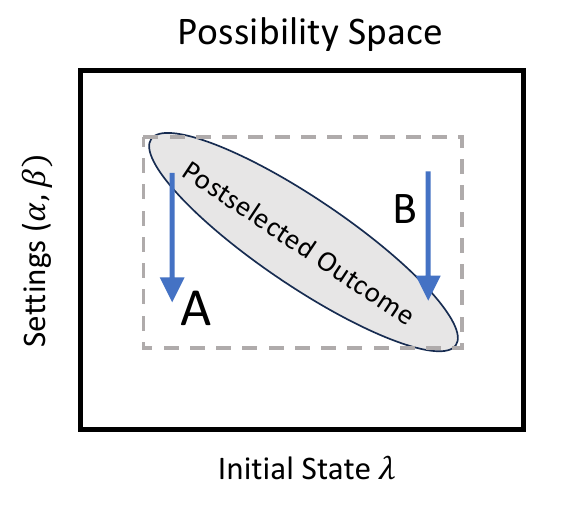}
\caption{This schematic indicates the generic correlation between settings and initial states in postselected subensembles.  In general a subensemble might be a probability distribution in possibility space, but here it is represented as a deterministic region (as in our classical model).  The vertical axis represents the multidimensional space of measurements one might make on a system, and the horizontal axis represents the multidimensional space of initial states $\lambda$, including any hidden/unknown variables. Counterfactual changes of the settings, keeping $\lambda$ fixed, can eliminate some values of $\lambda$ from the subensemble (arrow $A$) and/or bring new values into the subensemble (arrow $B$). Any attempt to remove this correlation by expanding the subensemble (dashed box) must include discarded runs or null-results.  }
\label{fig3}
\end{center}
\end{figure}

It is important to set aside one obvious method of postselecting this experiment, a method that is not being suggested in \cite{Wang}.  One \textit{could} define $Z$ to be rate at which at least one side of the experiment makes a two-photon detection.  In the classical analog model, for the settings with the highest four-photon rate ($N=4.6\%$), this turns out to be $Z=36.7\%$.  If one then discards all other runs (those for which neither side makes even a two-photon detection), this boosts the relative probability to $p=4.6/36.7=12.5\%$.  But this is not large enough for Bell-violations, and the experimental probabilities calculated in this way are much smaller than even $12.5\%$ (assuming one can compare the vertical scales in the various results presented in \cite{Wang}).

The postselection described in \cite{Wang} are cases where all four photons are measured at the end of the experiment, not merely two.  But then how should the measured rates be normalized?  For any given setting combination $(\alpha,\beta)$, the four-photon-coincidence rate is the very definition of $N(\alpha,\beta)$ in the first place.  Setting $Z$ to this same value would just lead to the trivial $p=1$.  Instead, the authors of \cite{Wang} appear to take it as evident that $Z$ should be the ``rate'' at which four-photon-coincidences \textit{could have been} detected, measured by varying both angles by $\pm\pi$ and adding those four rates together.  But this enters the domain of counterfactual reasoning, a subtle issue for any postselected experiment.  This topic will now be addressed in detail before returning to the experiment in question.

\subsection {Counterfactuals and Bell's Theorem}
``Counterfactual'' is  a technical term for a simple and common concept.  When using a physics model, it is natural to ask what would be different if some inputs to the model were changed while other inputs stayed the same.  When students first encounter $F=ma$, they often are asked to consider the effects of different forces while holding the mass constant.  In any particular physical instance, there can only be one actual force, but this simple model can be used in such a ``counterfactual'' manner all the same.\footnote{Judea Pearl \cite{Causality,Why} has argued that this is the foundational basis for essentially all causal reasoning, and our minds perform this procedure so naturally we sometimes hardly even notice.}  

J.~S.~Bell also discussed counterfactuals in these terms, most explicitly when he was setting aside so-called ``superdeterministic'' objections \cite{bell1977}.  What mattered, as Bell saw it, was calculating the effect of different externally-imposed measurement settings while holding the relevant part of the past constant (in present terminology, fixing $\lambda$ while changing $(\alpha,\beta)$).  The central question of Bell's analysis was: What would have been the (counterfactual) outcome with the same $\lambda$ but different measurement settings $(\alpha,\beta)$?  And when formalizing his results, he found that his inequalities were only derivable when imposing a ``Statistical Independence'' (SI) between $\lambda$ and  $(\alpha,\beta)$:
\begin{equation}
\label{eq:SI}
    P(\lambda|\alpha,\beta)=P(\lambda).
\end{equation}
In other words, if SI is violated, one would generally expect Bell inequalities to be violated as well, even in a classical model without any action-at-a-distance.\footnote{This has been explicitly demonstrated in a number of such models \cite{WhartonArgaman}.}

Whatever one thinks of this SI condition in the full possibility space of a given physics model, it is incontrovertible that SI can be broken by postselecting some portion of that space, as shown in Figure 3.  Such a correlation is almost inevitable, if the outcome to an experiment is jointly caused by the initial state $\lambda$ and the measurement settings. In causal modeling it is well known that postselecting on a jointly-caused outcome will typically induce correlations between the causes \cite{Causality}, as described in the Introduction. Given such a correlation, as schematically shown in Figure 3, Bell inequality violations would be permitted in the postselected subensemble. 

The reason for these violations is the counterfactual fragility \cite{PriceWharton21a} of Bell-style counterfactual reasoning in a subensemble.  Given a measurement $N(\alpha,\beta)$ for some particular setting, the standard counterfactual question is simply: if the settings had been different, but $\lambda$ had been the same, what would be the result?  But this change of settings may throw the experiment out of the subensemble, corresponding to arrow $A$ in Figure 3. Within the constraints of the postselected subensemble, then, it is not possible for the settings to be freely chosen while fixing $\lambda$.

There is a different form of counterfactual reasoning we might employ, which would allow a change of settings. It is a so-called  \textit{backtracking} counterfactual \cite{lewis79,vonk23}.  In general, this is when we consider a non-actual possibility in the present, and ``look backwards'', asking ourselves what the history \textit{would have been,} in that situation. (Luckily our nuclear reactor did not explode today, but if \textit{had} exploded, what would have caused the explosion? What is the most likely counterfactual history, to produce that result?)

In the present case, then, we can ask: Suppose we had measured an actual result (or at least a result in our postselected subensemble), but with different settings $(\alpha,\beta)$. What would the history have been like, in that case? Here, clearly, we do not ``hold the past fixed'', and the answer will be that $\lambda$ would have been different, in general. We can get into the postselected subensemble with different settings, but in general we need to start with a different initial state $\lambda$ (and from there, changing the settings as indicated by arrow $B$ in Figure 3).  The net effect is that we get correlations between the settings and $\lambda$ in the subensemble. This isn't surprising, as the relevant region in Figure 3 should make clear. It is just collider bias at work. 

If there were to be any hope of having a robust Bell-style no-go theorem in a postselected subensemble, one would have to remove the correlation between the settings and $\lambda$.  This might be done by ``postselecting'' a larger ensemble -- such as the dashed box shown in Figure 3, enclosing the entire shaded region.  But now, ``postselecting'' needs to be in scare-quotes because there is no actual \textit{selection} one can perform outside of the original region.  This larger region now necessarily includes null-results, corresponding to a detection rate of zero (given a particular value of $\lambda$).  The technique that \cite{Wang} proposes to incorporate such null-results now requires further discussion.  

\subsection{Counterfactuals and Null Results}\

The authors of \cite{Wang} indicate that a template for considering null-results is the case of a single photon measured by an absorbing polarizer and a single detector.  In this case (presumably given a single-photon source and a perfect detector) the authors note that counterfactual reasoning about a null result seems reasonable.  Their paper states:

\begin{quote}
Specifically, when the polarizer is set to $0^\circ$, only the $+1$ (H) outcome is detected, while the orthogonal $-1$ (V) outcome is blocked. These experiments typically rely on the assumption based on Malus's law, the cosine dependence of the intensity of a polarized beam after an ideal polarizer. This assumption permits one to use the $+1$ outcome measured with the polarizer oriented at $90^\circ$ to infer the otherwise inaccessible $-1$ outcome at $0^\circ$. In other words, by performing selective projective measurements along these mutually orthogonal directions, the combined outcomes form a complete projective measurement.
\end{quote}

The idea here is that values of $\lambda$ that would lead to a null result for one measurement setting $M_1$ might instead be measured by a different measurement setting $M_2$.  Together, these two measurements would form a ``complementary set'', in principle being able to collectively sample all values of $\lambda$ in some desired region of possibility space.  

More generally, the argument would go like this.  Suppose a source produces initial states with a probability distribution $p_0(\lambda)$, independent of the future measurement setting.  As discussed above, the probability of measuring a postselected outcome $p_{ps}(\lambda,M_i)$ with a measurement setting $M_i$ will almost always be different from $p_0(\lambda)$, breaking Statistical Independence.\footnote{This is not simply a matter of changing the overall normalization; some values of $\lambda$ might be completely unrepresented in the postselected subensemble.} Nevertheless, one might hope to find two sets of $N$ measurement settings, $\{M_1,M_2,...,M_N\}$ and $\{M'_1,M'_2,...,M'_N\}$, such that
\begin{equation}
\label{eq:cset}
    \sum_{i=1}^N p_{ps}(\lambda,M_i)=\sum_{i=1}^N p'_{ps}(\lambda,M'_i)=p_{ps}(\lambda),
\end{equation}
where $p_{ps}(\lambda)$ was some representative subset of initial states that could stand in for $p_0(\lambda)$ in a proof of Bell's Theorem.  By thinking of the entire set $M_i$ or $M'_i$ as a choice of ``setting'', an effective sort of Statistical Independence might be recovered.

This definition would appear to work as they describe for a single-photon polarization measurement.  A complementary set of two measurements would consist of a polarization measurement $M_1$ at the angle $\alpha$ and another polarization measurement $M_2$ at the angle $\alpha+\pi/2$.  (Every choice of $\alpha$ would then yield a different complementary set.)  Given a heralded single photon prepared with some polarization $\theta$ -- or, more generally, a probability distribution $p_0(\theta)$ -- Malus's Law would predict $p_{ps}(\theta,M_1)=\cos^2(\theta-\alpha)p_0(\theta)$ and $p_{ps}(\theta,M_2)=\sin^2(\theta-\alpha)p_0(\theta)$.  For every possible value of $\alpha$, one finds $\sum p_{ps}(\theta,M_i)=p_0(\theta)$.  Interpreting $\theta$ as the hidden variable $\lambda$, Eqn. (\ref{eq:cset}) holds true.

However, it is often impossible to test Eqn. (\ref{eq:cset}) empirically.  Such a condition is model-dependent.  For this special heralded single-photon case, if one believes that one can swap out the absorbing polarizer for a polarizing beamsplitter without consequence, a second detector can confirm the success of this particular complementary set.  But in general, without control of $\lambda$, there is no way to experimentally test that $\sum p_{ps}(\lambda,M_i)=\sum p'_{ps}(\lambda,M'_i)$.  All that can be empirically observed is the marginal probability\footnote{Or worse, a mere postselection rate, requiring an additional normalization.} of a postselection for each setting,
\begin{equation}
    p_{ps}(M_i) = \sum_\lambda p_{ps}(\lambda,M_i)p_0(\lambda).
\end{equation}

This new version of Statistical Independence, as defined by Eqn. (\ref{eq:cset}), is therefore very easy to violate.  For instance, consider a value of $\lambda$ that generated some postselected outcomes for at least one setting in a proposed complementary set $M_i$, but then failed to generate any postselected outcomes for \textit{any} setting of a different set $M'_i$.  A different value of $\lambda$ could then have exactly the opposite behavior, hiding this effect from the marginal detection rates.  Changing the set and postselecting would then lead to two different distributions $p_{ps}(\lambda)$ and $p'_{ps}(\lambda)$, and neither would be a fair sample of the possible $\lambda$'s in general.  

Given this unfair-sampling scenario, changing the measurement set from $M_i$ to $M'_i$, would then simply alter the postselected $\lambda$ subspace that one was measuring.  This is exactly the same problem as described in the previous subsection, and the associated violation of Statistical Independence would void any attempt at a Bell-style no-go theorem.  Without any clear way to diagnose whether this is happening, using counterfactual reasoning in \textit{any} postselected system that includes null-results is deeply problematic.  Very few robust conclusions could be drawn from such an analysis.

\subsection{Comparison with the classical model}

In the previous section we saw that a simple classical model (inspired by the quantum experiment in \cite{Wang}) can generate the same observed postselected rates as the actual experiment.  And yet \cite{Wang} also argues that a Bell inequality is violated by these results, which some might think would be impossible to find in a classical model.  In this section we have seen the general explanation for this mismatch: the postselection acts to break the Statistical Indepdendence assumption on which Bell-type no-go theorems rely. 

But a more detailed comparison is possible in this case, because the classical model is so similar to the quantum experiments.  In \cite{Wang} the authors propose a family of complementary sets, each given by four possible setting combinations:
\begin{eqnarray}
    M_1 &=& (\alpha,\beta)\\
    M_2 &=& (\alpha+\pi,\beta)\\
    M_3 &=& (\alpha,\beta+\pi)\\
    M_4 &=& (\alpha+\pi,\beta+\pi)
\end{eqnarray}
For a given value of $(\alpha,\beta)$, one can apply these measurements to the deterministic classical model, and see if the generalized version of Statistical Independence holds, from Eqn. (\ref{eq:cset}).  

Consider two of the sets used in the actual experiment: four settings $M_i$ generated using $(\alpha=0,\beta=\pi/4)$, and four more settings $M'_i$ generated using $(\alpha=0,\beta=3\pi/4)$.  Each of these eight settings can be used to compute a probability of a successful postselection in the classical model, and -- more importantly -- the values of $\lambda$ that generate those outcomes can be saved and analyzed.  Running the classical simulation reveals that for the values of $\lambda$ for which $M_1$ was postselected, only about half of those $\lambda$'s also lead to a postselection in the entire set $M'_i$.   

This means that roughly half of the meaningful space of $\lambda$'s in one set, $M_i$, corresponds to $\sum p'_{ps}=0$ in the other set $M'_i$.  The converse is true as well, meaning that neither complementary set fairly samples the space of possible $\lambda$'s.  Shifting $\beta$ by $\pi/2$ completely changes the space of $\lambda$ that is being postselected by this experiment.  As discussed above, this violates Statistical Independence, and in turn explains the ``violation'' of the inequality in the classical model.  Presumably the same explanation could also be applied to the quantum experiment described in \cite{Wang}.

\section{Conclusions}

Wang \textit{et al} \cite{Wang} have reported a violation of a Bell inequality without entanglement.  We have not taken a position on whether there might be hidden entanglement in their experiment, but have instead shown that there is no fundamental \textit{need} for entanglement to explain their results.  The strongest evidence for our claim is the classical model summarized in Section III, a Locally Causal model that generates essentially the same probabilities as their experiment.  Notably, this model was motivated by a classical analysis of their precise experimental geometry, as discussed in Section II.  

In the context of our classical hidden variable model, Section IV showed precisely how the Bell inequality violation is achieved.  In the postselected subensemble of model results, we find a clear correlation between the settings $(\alpha,\beta)$ and the initial hidden variables $(\delta_1,\delta_2,\delta_3,\delta_4)$.  Even when expanding the analysis to complementary sets of measurements, rather than single measurement settings, this correlation persists.  And any such violation of Statistical Independence is known to generally allow for Bell inequality violations \cite{WhartonArgaman}.  This violation is easy to understand in causal modeling terms. It is collider bias, or Berkson's bias.

\section*{Acknowledgements}

The authors gratefully thank N. Argaman for helpful comments and discussion.

\newpage

\bibliography{References}

\end{document}